\documentclass[a4paper,11pt]{article}
\usepackage{graphicx}
\newcommand{\siki}[1]{Eq.\ref{eq:#1}}
\newcommand{\zu}[1]{Fig.\ref{fig:#1}}

\newcommand{\rref}[1]{${}^{\mathrm{#1)}}$}
\newcommand{\ep}{\varepsilon}

\begin{document}
%----------------------------------------------
\title{A note on the dimensional regularization and the on-mass-shell
renormalization in the two-loop order\footnote{%
Kogakuin Daigaku Kenkyuronso, vol.60-2, pp.1-8, 2023,  
https://doi.org/10.57377/00000342 }} 
\author{Kiyoshi KATO\footnote{e-mail\ :\ {\tt katok@kute.tokyo}} \\
{ \small Kogakuin University, Nishi-Shinjuku 1-24, Shinjuku, Tokyo 160-0023, Japan} 
}
\date{}

\maketitle

%\begin{flushright}
%{\tt 2022ronsou.pdf} \\
%\end{flushright}

%\vspace{5mm}

%-----------------------------------------------
\begin{quote}
\noindent{\bfseries{Abstract}}\ 
The use of the dimensional regularization in the on-mass-shell renormalization scheme
sometimes fails to locally cancel the ultraviolet divergence for a class of diagrams
in the two-loop order.
The mechanism is discussed based on an example with explicit computation.
\end{quote}

%-----------------------------------------------
\section{Introduction}

The purpose of this note is to describe a phenomenon which appears in the 
calculation of Feynman diagrams beyond the one-loop order.
Though the content is elementary, its explicit explanation is seldom found in textbooks nor
literatures, so that it can be worth to record as a simple memorandum.

The problem discussed in this note is that sometimes the local cancelation of the ultraviolet(UV) divergence
seems to be incomplete in the two-loop order.

In this note, a few Feynman diagrams for the self-energy function of a scalar field $\phi$ 
are studied as an explicit example to describe the problem mentioned above.
While the scalar field $\phi$ interacts with other fields like the Higgs particle in the standard model,
the detailed structure of the interaction is not necessary in the discussion.

The renormalization is performed based on the on-mass-shell scheme.\rref{1}
The $\phi$ field looks like the Higgs, but no tadpole contribution is considered.

For the regularization of the UV divergence, the dimensional
regularization is used where the space-time dimension $n$ is given by
\begin{equation}
n=4 - 2\ep\ .
\end{equation}
Here $\ep$ is  an infinitesimal quantity and 
the UV divergence is no more abstract infinity but appears as the inverse of $\ep$,
so that the infinity is now under control in the computation. The renormalization
process is the cancelation of such divergent terms.

In the following sections, the analytic calculation of one- and two-loop diagrams
is performed. As it is usually hard to calculate two-loop diagrams exactly,
the leading UV singularity is mostly studied.  We sometimes call the UV
singularity in the form of ${\displaystyle \frac{1}{\ep}}$ and that of ${\displaystyle \frac{1}{\ep^2}}$
as the single pole and the double pole, respectively.

%-----------------------------------------------
\section{one-loop two-point function of $\phi$ and its counter term}

The one-loop contribution to the two-point function of $\phi$ is given by the diagrams in \zu{figone}.
The external momentum $q$ enters in the diagram. 
In this note, the propagator of $\phi$ is defined as $1/(p^2-M^2)$.

\begin{figure}[htb]
\begin{center}
%WinTpicVersion4.28b
{\unitlength 0.1in
\begin{picture}( 27.9000, 10.3000)(  4.5000,-12.5000)
% ELLIPSE 2 0 3 0 Black White
% 4 2400 800 3000 1200 3000 1200 3000 1200
% 
\special{pn 8}%
\special{ar 2400 800 600 400  0.0000000  6.2831853}%
% DOT 0 0 3 0 Black White
% 3 3000 800 1800 800 1800 800
% 
\special{pn 4}%
\special{sh 1}%
\special{ar 3000 800 16 16 0  6.28318530717959E+0000}%
\special{sh 1}%
\special{ar 1800 800 16 16 0  6.28318530717959E+0000}%
\special{sh 1}%
\special{ar 1800 800 16 16 0  6.28318530717959E+0000}%
% VECTOR 2 0 3 0 Black White
% 2 1600 800 1730 800
% 
\special{pn 8}%
\special{pa 1600 800}%
\special{pa 1730 800}%
\special{fp}%
\special{sh 1}%
\special{pa 1730 800}%
\special{pa 1664 780}%
\special{pa 1678 800}%
\special{pa 1664 820}%
\special{pa 1730 800}%
\special{fp}%
% VECTOR 2 0 3 0 Black White
% 2 3110 800 3240 800
% 
\special{pn 8}%
\special{pa 3110 800}%
\special{pa 3240 800}%
\special{fp}%
\special{sh 1}%
\special{pa 3240 800}%
\special{pa 3174 780}%
\special{pa 3188 800}%
\special{pa 3174 820}%
\special{pa 3240 800}%
\special{fp}%
% STR 2 0 3 0 Black White
% 4 1500 640 1500 740 2 0 0 0
% $q$
\put(15.0000,-7.4000){\makebox(0,0)[lb]{$q$}}%
% STR 2 0 3 0 Black White
% 4 3230 640 3230 740 2 0 0 0
% $q$
\put(32.3000,-7.4000){\makebox(0,0)[lb]{$q$}}%
% STR 2 0 3 0 Black White
% 4 2180 250 2180 350 2 0 0 0
% $\ell+q, \ x$
\put(21.8000,-3.5000){\makebox(0,0)[lb]{$\ell+q, \ x$}}%
% STR 2 0 3 0 Black White
% 4 2140 1280 2140 1380 2 0 0 0
% $\ell, \ 1-x$
\put(21.4000,-13.8000){\makebox(0,0)[lb]{$\ell, \ 1-x$}}%
% STR 2 0 3 0 Black White
% 4 2200 500 2200 600 2 0 0 0
% 1,\ $m_1$
\put(22.0000,-6.0000){\makebox(0,0)[lb]{1,\ $m_1$}}%
% STR 2 0 3 0 Black White
% 4 2190 1030 2190 1130 2 0 0 0
% 2,\ $m_2$
\put(21.9000,-11.3000){\makebox(0,0)[lb]{2,\ $m_2$}}%
% VECTOR 2 0 3 0 Black White
% 2 2400 400 2430 400
% 
\special{pn 8}%
\special{pa 2400 400}%
\special{pa 2430 400}%
\special{fp}%
\special{sh 1}%
\special{pa 2430 400}%
\special{pa 2364 380}%
\special{pa 2378 400}%
\special{pa 2364 420}%
\special{pa 2430 400}%
\special{fp}%
% VECTOR 2 0 3 0 Black White
% 2 2356 1200 2326 1200
% 
\special{pn 8}%
\special{pa 2356 1200}%
\special{pa 2326 1200}%
\special{fp}%
\special{sh 1}%
\special{pa 2326 1200}%
\special{pa 2394 1220}%
\special{pa 2380 1200}%
\special{pa 2394 1180}%
\special{pa 2326 1200}%
\special{fp}%
% STR 2 0 3 0 Black White
% 4 450 770 450 870 2 0 0 0
% $\Pi_{loop}(q^2)=$
\put(4.5000,-8.7000){\makebox(0,0)[lb]{$\Pi_{loop}(q^2)=$}}%
\end{picture}}%
\end{center}
\caption{one-loop two-point function of $\phi$ }
\label{fig:figone}
\end{figure}

\begin{equation}
\Pi_{loop}(q^2)=\int[d\ell] \frac{N}{D_1D_2} ,\quad
D_1= (\ell+q)^2-m_1^2 , \quad D_2=\ell^2-m_2^2
\label{eq:mainint}
\end{equation}
where $[d\ell]$ is defined in the appendix-2 and 
the numerator $N$ is given by 
\begin{equation}
N= a + b(q^2) + c(q\ell) + d(\ell^2) \ .
\end{equation}
Here $a, b, c, d$ are constants which are determined by the particles in the loop
and the interaction.  The numerator structure is sufficiently general if
we follow the standard model with Feynman gauge.

\siki{mainint} is calculated using formulae given in the appendix-2.
\begin{equation}
\frac{1}{D_1D_2}= \int_0^1 dx \frac{1}{[xD_1 +(1-x)D_2]^2} .
\end{equation}
After the momentum shift $\ell \rightarrow \ell-xq$ and omitting odd terms in $\ell$,
\begin{equation}
\Pi_{loop}(q^2)=\int_0^1 dx \int[d\ell] \frac{N_1}{[\ell^2-D_{12}]^2}
\end{equation}
where
\begin{equation}
D_{12}=m_1^2 x + m_2^2 (1-x) - q^2 x(1-x), \qquad
N_1 =  [a + (b-cx+dx^2) q^2] + d \ell^2 \ .
\end{equation}

We perform the $\ell$-integration and obtain
\begin{equation}
\Pi_{loop}(q^2)=(FAC_1) \frac{1}{\ep} \int_0^1 dx \left[
  e + f q^2
\right]\frac{1}{D_{12}^{\ep}(q^2)}
\end{equation}
where $(FAC_1)=\Gamma(1+\ep)(4\pi)^{\ep}/(16\pi^2)$ and
\begin{equation}
e = a + d \frac{2-\ep}{1-\ep}(m_1^2 x + m_2^2 (1-x)) , \quad
f = (b-cx+dx^2) - d \frac{2-\ep}{1-\ep} x(1-x) .
\end{equation}

The $x$-integrals of $e$ and $f$ are
\begin{equation}
\int_0^1 dx e = \bar{e}+ O(\ep) = a + d(m_1^2+m_2^2) + O(\ep) ,\qquad
\int_0^1 dx f = \bar{f}+ O(\ep) = b-\frac{1}{2}c + O(\ep) \ .
\end{equation}

The derivative of the function is as follows.
\begin{equation}
\Pi_{loop}'(q^2)=(FAC_1)\int_0^1 dx \left[ \frac{1}{\ep} f + \frac{(e+ fq^2)x(1-x)}{D_{12}(q^2)}\ .
\right]\frac{1}{D_{12}^{\ep}(q^2)}
\end{equation}

The renormalized mass of $\phi$-field is denoted by $M$.
From the renormalization condition in the on-mass-shell scheme, the counter term is
give by\rref{1} 
\begin{equation}
\Pi_{ct}(q^2) = A(q^2-M^2) + B, \qquad 
A=-\Pi_{loop}'(M^2), \quad B= - \Pi_{loop}(M^2) \ 
\label{eq:defab}
\end{equation}
where 
\begin{equation}
A= -\frac{1}{\ep} (FAC_1)  \bar{f}  + O(1) ,\quad
B= -\frac{1}{\ep} (FAC_1)  ( \bar{e} + \bar{f} M^2 ) + O(1) \ .
\label{eq:defabx}
\end{equation}
Here, $A$ is the wave function renormalization constant and 
$B$ is the mass renormalization constant.
It should be noted that the divergence of constants is determined after fixing
$q^2=M^2$.

Since the renormalization is performed, the following sum is finite at any $q^2$.
\begin{equation}
\Pi_{loop}(q^2)\ + \ \Pi_{ct}(q^2) 
= \ \mathrm{finite} \left( \ \mathrm{no} \ \frac{1}{\ep} \ \mathrm{term} \ \right)
\label{eq:sumpi}
\end{equation}

%\newpage
%-----------------------------------------------
\section{two-loop two-point function of $\phi$} 

Now we consider the diagrams in \zu{figtwo} and study the sum of them.
Naive expectation is as follows: While generally a two-loop diagram
can involve the double-pole singularity, the sum shall cancel such terms and
the remaining divergence shall be  the single-pole type due to the $\ell$ integration in the figure.

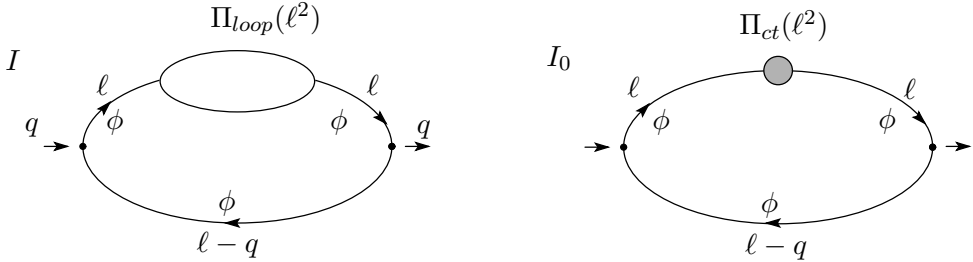
\begin{figure}[htb]
\begin{center}
%WinTpicVersion4.28b
{\unitlength 0.1in
\begin{picture}( 49.9000, 11.7000)(  2.0000,-14.6000)
% ELLIPSE 2 0 3 0 Black White
% 4 1400 1000 2200 1400 1000 660 1800 660
% 
\special{pn 8}%
\special{ar 1400 1000 800 400  5.2441130  4.1806649}%
% DOT 0 0 3 0 Black White
% 3 600 1000 2200 1000 2200 1000
% 
\special{pn 4}%
\special{sh 1}%
\special{ar 600 1000 16 16 0  6.28318530717959E+0000}%
\special{sh 1}%
\special{ar 2200 1000 16 16 0  6.28318530717959E+0000}%
\special{sh 1}%
\special{ar 2200 1000 16 16 0  6.28318530717959E+0000}%
% ELLIPSE 2 0 3 0 Black White
% 4 4200 1003 5000 1403 5000 1403 5000 1403
% 
\special{pn 8}%
\special{ar 4200 1004 800 400  0.0000000  6.2831853}%
% DOT 0 0 3 0 Black White
% 3 3400 1003 5000 1003 5000 1003
% 
\special{pn 4}%
\special{sh 1}%
\special{ar 3400 1004 16 16 0  6.28318530717959E+0000}%
\special{sh 1}%
\special{ar 5000 1004 16 16 0  6.28318530717959E+0000}%
\special{sh 1}%
\special{ar 5000 1004 16 16 0  6.28318530717959E+0000}%
% CIRCLE 2 0 1 0 Black Black
% 4 4200 603 4220 673 4220 673 4220 673
% 
\special{sh 0.300}%
\special{ia 4200 604 74 74  0.0000000  6.2831853}%
\special{pn 8}%
\special{ar 4200 604 74 74  0.0000000  6.2831853}%
% VECTOR 2 0 3 0 Black White
% 2 3200 1003 3320 1003
% 
\special{pn 8}%
\special{pa 3200 1004}%
\special{pa 3320 1004}%
\special{fp}%
\special{sh 1}%
\special{pa 3320 1004}%
\special{pa 3254 984}%
\special{pa 3268 1004}%
\special{pa 3254 1024}%
\special{pa 3320 1004}%
\special{fp}%
% VECTOR 2 0 3 0 Black White
% 2 400 1000 520 1000
% 
\special{pn 8}%
\special{pa 400 1000}%
\special{pa 520 1000}%
\special{fp}%
\special{sh 1}%
\special{pa 520 1000}%
\special{pa 454 980}%
\special{pa 468 1000}%
\special{pa 454 1020}%
\special{pa 520 1000}%
\special{fp}%
% VECTOR 2 0 3 0 Black White
% 2 5070 1000 5190 1000
% 
\special{pn 8}%
\special{pa 5070 1000}%
\special{pa 5190 1000}%
\special{fp}%
\special{sh 1}%
\special{pa 5190 1000}%
\special{pa 5124 980}%
\special{pa 5138 1000}%
\special{pa 5124 1020}%
\special{pa 5190 1000}%
\special{fp}%
% VECTOR 2 0 3 0 Black White
% 2 2270 1000 2390 1000
% 
\special{pn 8}%
\special{pa 2270 1000}%
\special{pa 2390 1000}%
\special{fp}%
\special{sh 1}%
\special{pa 2390 1000}%
\special{pa 2324 980}%
\special{pa 2338 1000}%
\special{pa 2324 1020}%
\special{pa 2390 1000}%
\special{fp}%
% ELLIPSE 2 0 3 0 Black White
% 4 1400 660 1800 820 1800 820 1800 820
% 
\special{pn 8}%
\special{ar 1400 660 400 160  0.0000000  6.2831853}%
% VECTOR 2 0 3 0 Black White
% 2 1400 1400 1340 1400
% 
\special{pn 8}%
\special{pa 1400 1400}%
\special{pa 1340 1400}%
\special{fp}%
\special{sh 1}%
\special{pa 1340 1400}%
\special{pa 1408 1420}%
\special{pa 1394 1400}%
\special{pa 1408 1380}%
\special{pa 1340 1400}%
\special{fp}%
% VECTOR 2 0 3 0 Black White
% 2 700 810 740 770
% 
\special{pn 8}%
\special{pa 700 810}%
\special{pa 740 770}%
\special{fp}%
\special{sh 1}%
\special{pa 740 770}%
\special{pa 680 804}%
\special{pa 702 808}%
\special{pa 708 832}%
\special{pa 740 770}%
\special{fp}%
% VECTOR 2 0 3 0 Black White
% 2 2130 840 2170 880
% 
\special{pn 8}%
\special{pa 2130 840}%
\special{pa 2170 880}%
\special{fp}%
\special{sh 1}%
\special{pa 2170 880}%
\special{pa 2138 820}%
\special{pa 2132 842}%
\special{pa 2110 848}%
\special{pa 2170 880}%
\special{fp}%
% STR 2 0 3 0 Black White
% 4 670 630 670 730 2 0 0 0
% $\ell$
\put(6.7000,-7.3000){\makebox(0,0)[lb]{$\ell$}}%
% STR 2 0 3 0 Black White
% 4 2070 630 2070 730 2 0 0 0
% $\ell$
\put(20.7000,-7.3000){\makebox(0,0)[lb]{$\ell$}}%
% STR 2 0 3 0 Black White
% 4 1190 1480 1190 1580 2 0 0 0
% $\ell-q$
\put(11.9000,-15.8000){\makebox(0,0)[lb]{$\ell-q$}}%
% STR 2 0 3 0 Black White
% 4 720 840 720 940 2 0 0 0
% $\phi$
\put(7.2000,-9.4000){\makebox(0,0)[lb]{$\phi$}}%
% STR 2 0 3 0 Black White
% 4 1260 320 1260 420 2 0 0 0
% $\Pi_{loop}(\ell^2)$
\put(12.6000,-4.2000){\makebox(0,0)[lb]{$\Pi_{loop}(\ell^2)$}}%
% STR 2 0 3 0 Black White
% 4 4000 350 4000 450 2 0 0 0
% $\Pi_{ct}(\ell^2)$
\put(40.0000,-4.5000){\makebox(0,0)[lb]{$\Pi_{ct}(\ell^2)$}}%
% STR 2 0 3 0 Black White
% 4 300 850 300 950 2 0 0 0
% $q$
\put(3.0000,-9.5000){\makebox(0,0)[lb]{$q$}}%
% STR 2 0 3 0 Black White
% 4 2330 860 2330 960 2 0 0 0
% $q$
\put(23.3000,-9.6000){\makebox(0,0)[lb]{$q$}}%
% STR 2 0 3 0 Black White
% 4 1880 840 1880 940 2 0 0 0
% $\phi$
\put(18.8000,-9.4000){\makebox(0,0)[lb]{$\phi$}}%
% STR 2 0 3 0 Black White
% 4 1300 1260 1300 1360 2 0 0 0
% $\phi$
\put(13.0000,-13.6000){\makebox(0,0)[lb]{$\phi$}}%
% STR 2 0 3 0 Black White
% 4 3550 840 3550 940 2 0 0 0
% $\phi$
\put(35.5000,-9.4000){\makebox(0,0)[lb]{$\phi$}}%
% STR 2 0 3 0 Black White
% 4 4110 1270 4110 1370 2 0 0 0
% $\phi$
\put(41.1000,-13.7000){\makebox(0,0)[lb]{$\phi$}}%
% STR 2 0 3 0 Black White
% 4 4720 830 4720 930 2 0 0 0
% $\phi$
\put(47.2000,-9.3000){\makebox(0,0)[lb]{$\phi$}}%
% STR 2 0 3 0 Black White
% 4 200 500 200 600 2 0 0 0
% $I$
\put(2.0000,-6.0000){\makebox(0,0)[lb]{$I$}}%
% STR 2 0 3 0 Black White
% 4 3000 500 3000 600 2 0 0 0
% $I_0$
\put(30.0000,-6.0000){\makebox(0,0)[lb]{$I_0$}}%
% STR 2 0 3 0 Black White
% 4 4030 1490 4030 1590 2 0 0 0
% $\ell-q$
\put(40.3000,-15.9000){\makebox(0,0)[lb]{$\ell-q$}}%
% STR 2 0 3 0 Black White
% 4 3420 640 3420 740 2 0 0 0
% $\ell$
\put(34.2000,-7.4000){\makebox(0,0)[lb]{$\ell$}}%
% STR 2 0 3 0 Black White
% 4 4850 650 4850 750 2 0 0 0
% $\ell$
\put(48.5000,-7.5000){\makebox(0,0)[lb]{$\ell$}}%
% VECTOR 2 0 3 0 Black White
% 2 4200 1400 4140 1400
% 
\special{pn 8}%
\special{pa 4200 1400}%
\special{pa 4140 1400}%
\special{fp}%
\special{sh 1}%
\special{pa 4140 1400}%
\special{pa 4208 1420}%
\special{pa 4194 1400}%
\special{pa 4208 1380}%
\special{pa 4140 1400}%
\special{fp}%
% VECTOR 2 0 3 0 Black White
% 2 3490 820 3530 780
% 
\special{pn 8}%
\special{pa 3490 820}%
\special{pa 3530 780}%
\special{fp}%
\special{sh 1}%
\special{pa 3530 780}%
\special{pa 3470 814}%
\special{pa 3492 818}%
\special{pa 3498 842}%
\special{pa 3530 780}%
\special{fp}%
% VECTOR 2 0 3 0 Black White
% 2 4920 840 4960 880
% 
\special{pn 8}%
\special{pa 4920 840}%
\special{pa 4960 880}%
\special{fp}%
\special{sh 1}%
\special{pa 4960 880}%
\special{pa 4928 820}%
\special{pa 4922 842}%
\special{pa 4900 848}%
\special{pa 4960 880}%
\special{fp}%
\end{picture}}%
\end{center}
\caption{two-loop two-point function of $\phi$, $I$(left) and  $I_0$(right)}
\label{fig:figtwo}
\end{figure}

First, we calculate $I$ where $\ell_1$ and $\ell_2$ are the loop momentum of outer large loop
and that of inner small loop, respectively, and $N$ is the same in the last section 
since the numerator of $\phi$-propagator is $1$.
\begin{equation}
I(q^2)= \int [d\ell_1] [d\ell_2] \frac{N}{D_1D_2D_3D_4D_5}, 
\label{eq:twotwon}
\end{equation}
\begin{equation}
D_1= (\ell_2+\ell_1)^2-m_1^2 , \quad D_2=\ell_2^2-m_2^2, \quad
D_3=D_4=\ell_1^2-M^2, \quad D_5= (\ell_1-q)^2-M^2 .
\end{equation}

In the following, the successive integration is applied to the calculation of $I$.\rref{2}
The two-loop computation is done by $\ell_2$ integral first and  $\ell_1$ integral second
using formulae in the appendix-2.
The inner $\ell_2$-loop integral is already done in the last section to get $\Pi_{loop}$ and
hereafter we write the outer loop momentum $\ell_1$ as $\ell$ as shown in the figure.
\begin{equation}
I(q^2)= \int [d\ell] \frac{\Pi_{loop}(\ell^2)}{D_3D_4D_5}\ .
\end{equation}
Substituting the expression in the last section, we have
\begin{equation}
I(q^2)= (FAC_1) \frac{1}{\ep} \int_0^1 \frac{dx}{(-x(1-x))^{\ep}} \int [d\ell]
\frac{  e + f \ell^2 }{D_3^2 D_5 [\ell^2-M_X^2]^{\ep}} \ 
\end{equation}
where $D_{12}$ is written as 
\begin{equation}
D_{12}(\ell^2)=  -x(1-x) [ \ell^2 -M_X^2 ] , \qquad
M_X^2 = \frac{m_1^2x+m_2^2(1-x)}{x(1-x)} .
\end{equation}

The denominator is combined as 
\begin{equation}
\frac{1}{D_3^2 D_5 [\ell^2-M_X^2]^{\ep}} 
= \frac{\Gamma(3+\ep)}{\Gamma(1)\Gamma(2)\Gamma(\ep)}
\int_{u+v\le 1} dudv \frac{v(1-u-v)^{1-\ep}}{( uD_5 + vD_3 + (1-u-v) [\ell^2-M_X^2] )^{3+\ep} }.
\end{equation}
After the momentum shift $\ell \rightarrow \ell+uq$ and omitting odd terms in $\ell$,
we have the following $\ell$-integral
\begin{equation}
J=\int [d\ell] \frac{h_0+h_1 \ell^2}{(\ell^2-D)^{3+\ep}}
\end{equation}
where 
\begin{equation}
D(q^2)= M^2(u+v)+M_X^2(1-u-v)- q^2 u(1-u) , \quad
h_0= e + f u^2 q^2 , \quad
h_1 = f \ .
\end{equation}
The above $\ell$-integration is done to obtain 
\begin{equation}
J= \frac{(4\pi)^{\ep}}{16\pi^2}(-1)^{\ep}\left( 
 - \frac{\Gamma(1+2\ep)}{\Gamma(3+\ep)}h_0 \frac{1}{D^{1+2\ep}}
 + \frac{(2-\ep)\Gamma(2\ep)}{\Gamma(3+\ep)}h_1 \frac{1}{D^{2\ep}} 
\right)
\end{equation}
Finally we have
\begin{equation}
I(q^2)=(FAC_2) \int_0^1 dx \frac{1}{(x(1-x))^{\ep}}  \int_{u+v\le 1} dudv\, v(1-u-v)^{\ep-1}
\left( - h_0 \frac{1}{D} + \frac{n}{4\ep}h_1 
\right)\frac{1}{D^{2\ep}}
\label{eq:finforma}
\end{equation}
where  $(FAC_2)=\Gamma(1+2\ep)(4\pi)^{2\ep}/(16\pi^2)^2$.

Next, $I_0$ is calculated as a one-loop diagram.

\begin{equation}
I_0(q^2)= \int [d\ell] \frac{A(\ell^2-M^2) + B}{D_3D_4D_5}  
=  \int [d\ell]\left( \frac{A}{D_3 D_5} + \frac{B}{D_3^2 D_5} \right)
\end{equation}

The $\ell$-integration leads to
\begin{equation}
I_0(q^2)= (FAC_1) \int_0^1 dx \left[\frac{1}{\ep}A - (1-x) \frac{B}{D_0}\right] \frac{1}{D_0^{\ep}}
\label{eq:finformb}
\end{equation}
where
\begin{equation}
D_0(q^2)=M^2 - q^2 x(1-x) \ .
\end{equation}
Here the divergence $\ep^{-1}$ with $A$ originates from this  $\ell$-integration.

In the following, we compare $I(M^2)$ and $I_0(M^2)$.
The relation $(FAC_2)=(FAC_1)^2(1+ O(\ep^2))$ is noted here.

%------------------------
\subsection{single pole}

In this subsection, $b=c=d=0$ is assumed. Then,
\begin{equation}
e = \bar{e}=a, f=\bar{f}=0 , \qquad h_0=a, h_1=0 \ .
\end{equation}

From \siki{finforma}, $I(M^2)$ is
\begin{equation}
I(M^2)=(FAC_2) \int_0^1 dx \frac{1}{(x(1-x))^{\ep}}  \int_{u+v\le 1}^1 dudv\, v(1-u-v)^{\ep-1}
( -1) a \frac{1}{D(M^2)} \ \frac{1}{D^{2\ep}}
\end{equation}
and we evaluate this after replacing $u$ by  $z=1-u-v$
\begin{equation}
I(M^2)=(FAC_2)(-a) \int_0^1 dx \frac{1}{(x(1-x))^{\ep}}  \int_0^1 dv  \int_0^{1-v} dz\, v z^{\ep-1}
\frac{1}{D(M^2)} \ \frac{1}{D^{2\ep}} .
\end{equation}

We only interested in the contribution of $\ep^{-1}$. Then $D(M^2,z=0)=M^2(1-v+v^2)$ and 
\[
I(M^2)\simeq (FAC_2)(-a) \int_0^1 dx \frac{1}{(x(1-x))^{\ep}}  \int_0^1 dv  \int_0^{1-v} dz\, v z^{\ep-1}
\]
\begin{equation}
\left[ \frac{1}{D(M^2,z=0)} + \left( \frac{1}{D(M^2)} - \frac{1}{D(M^2,z=0)} \right)\right] .
\end{equation}
Only the first term has the  $O(\ep^{-1})$ contribution which results from $z$-integration, so that
\begin{equation}
I(M^2) = (FAC_2)(-a)   \int_0^1 dv  \frac{v}{M^2(1-v+v^2)}  \frac{1}{\ep} + O(1)
=  (FAC_2)(-a) \frac{1}{\ep} \frac{k_0}{2M^2} + O(1) \ ,
\end{equation}
where
\begin{equation}
k_0=\int_0^1 \frac{dx}{1-x+x^2} = \frac{2\sqrt{3}\pi}{9} \ .
\end{equation}

Next, $I_0$ is evaluated. Since $\bar{f}=0, \bar{e}=a$, \siki{finformb} is
\[
I_0(M^2)= (FAC_1) \int_0^1 dx \left[  \frac{1}{\ep}A + (1-x)(FAC_1)a \frac{1}{\ep} \frac{1}{D_0(M^2)} \right] +O(1)
\]
\begin{equation}
= \frac{1}{\ep} \left[(FAC_1) \int_0^1 dx A + (FAC_1)^2 a \frac{k_0}{2M^2} \right] + O(1) .
\end{equation}
Here $A$ is a finite term.

\vspace{4mm}
\noindent \underline{ Result-1 }

The single-pole divergence  cancels between $I$ and $B$-term contribution of $I_0$.
The sum $I+I_0$ has divergence of $O(\ep^{-1})$ and the divergence comes from the outer $\ell$-integral 
of $A$-term.
Since $A$-term has the structure of $A(\ell^2-M^2)$, it cancels one denominator in $\ell$-loop
to produce divergence.

%------------------------
\subsection{double pole}

In this subsection,  $b, c, d$ are not vanishing.
We only interested in the contribution of $O(\ep^{-2})$.

In \siki{finforma}, $O(\ep^{-2})$ contribution comes from $h_1$-term.
\begin{equation}
I(M^2)=(FAC_2) \int_0^1 dx \frac{1}{(x(1-x))^{\ep}}  \int_{u+v\le 1}^1 dudv\, v(1-u-v)^{\ep-1}
\frac{1}{\ep}h_1  + O(\ep^{-1})
\end{equation}
Here $uv$-integral gives ( as is in the last subsection )
\begin{equation}
\int_{u+v\le 1}^1 dudv\, v(1-u-v)^{\ep-1} = \frac{1}{2\ep}+O(1) \ .
\end{equation}
Then we have 
\begin{equation}
I(M^2) = (FAC_2) \frac{1}{2\ep^2} \bar{f}  + O(\ep^{-1}) \ .
\label{eq:dpolea}
\end{equation}

In \siki{finformb}, $O(\ep^{-2})$ contribution comes from $A$-term
\begin{equation}
I_0(M^2) = (FAC_1) \int_0^1 dx \frac{1}{\ep}A  + O(\ep^{-1})
= -(FAC_1)^2 \frac{1}{\ep^2} \bar{f}  + O(\ep^{-1})
\end{equation}

\vspace{4mm}
\noindent \underline{ Result-2 }

The leading double-pole divergence  dose not cancel between
$I$ and $I_0$. The factor 2 difference exists in the result.
This is  annoying as it differs from the expectation stated in the
beginning of this section.

%\newpage
%-----------------------------------------------
\section{discussion}

We have checked the results in the last section as follows.
The integral $I$  can be calculated 
by the standard Nakanishi formula\rref{3,4} and it is given in the appendix-1.
The obtained formula is examined analytically and it gives the same
leading UV divergence.
Complete analytic calculation of $I$ in the last section nor that in
the appendix can not be possible.  However, numerical treatment is
possible to evaluate the value of $I$. Especially, the direct computation method(DCM)\rref{5,6}
%%developed by the research group including the author 
is able to
compute the coefficients of Laurent expansion of $I$ in $\ep$. 
The value of the coefficients of $\ep^{-1}$($\ep^{-2}$) in the single(double) pole case 
agrees with analytical results in good numerical accuracy.\rref{7}
Thus the analytic evaluation in the last section is confirmed numerically.

Another method to check \siki{dpolea} is to expand the numerator of \siki{twotwon} as
\begin{equation}
N=\left(a+bM^2+\frac{c}{2}(m_1^2-m_2^2-M^2)+dm_2^2\right)+\frac{c}{2}D_1
+\left(d-\frac{c}{2}\right)D_2 
+\left(b-\frac{c}{2}\right)D_3 \ . 
\end{equation}
Then $I$ is described by the linear combination of 4 two-loop integrals
whose numerators are $1$. Among them only $D_3$ term gives $O(\ep^{-2})$ contribution,\rref{8} i.e.,
\begin{equation}
 \int [d\ell_1] [d\ell_2] \frac{1}{D_1D_2D_4D_5}=(FAC_2) \frac{1}{2\ep^2} + O(\ep^{-1})\ .
\end{equation}

The source of discrepancy in the result-2 can be traced through the calculation.
In section 2, the renormalization constants $A, B$ are calculated using the value
of $D_{12}(q^2)$ at $q^2=M^2$ and expanded in $\ep$ to extract the divergent component 
assuming that $\ep$ is infinitesimal. 
On the other hand, in the two-loop calculation $D_{12}$
is treated as $D_{12}(\ell^2)$ and in the integral $|\ell^2|$ extends to infinity.
In the derivation of \siki{finforma},  $(\ell^2)^{\ep}$ from $\Pi_{loop}$ 
contributes to a factor $\Gamma(2\ep)$ in the integral $J$.
An inadvertent interchange of the two limiting processes,
$\ep\rightarrow 0$ and $|\ell^2|\rightarrow \infty$, seems to result the discrepancy.

For the regularization of UV divergence, we use the dimensional regularization in this note.
When we switch to the Pauli-Villars regularization\rref{9}
in the double pole case, the coefficients of $(\log \Lambda^2)^2$
cancels each other in the sum of $I+I_0$. Here $\Lambda^2$ is the large cutoff
parameter in the method.
This suggests that the problem presented in this note is related to
the dimensional regularization.

A possible modification to avoid the problem is as follows.
Though \siki{defabx} is the standard formula for $A, B$, the mass-dimension differs from \siki{defab},
so that a modified counter term can be proposed as 
\begin{equation}
\Pi_{ct}(q^2) =[ A(q^2-M^2) + B ] (M^2-q^2)^{-\ep} \ .
\end{equation}
When we take the limit $\ep\rightarrow 0$, this formula
is the same as the conventional one.
Then we have
\begin{equation}
I_0(q^2)= \int [d\ell] \frac{[ A(\ell^2-M^2) + B] (M^2-q^2)^{-\ep}}{[\ell^2-M^2]^2[(\ell-q)^2-M^2]} .
\end{equation}
After some manipulation, we obtain the following.
\begin{equation}
I_0(q^2)= (FAC_1) \frac{\Gamma(1+2\ep)}{\Gamma^2(1+\ep)} \int_0^1 dx \left[
\frac{1}{2\ep}A x^{\ep} -   x^{1+\ep} \frac{B}{(1+\ep)D_0}
\right] \frac{1}{D_0^{2\ep}}
\end{equation}
When we use the modified $I_0(M^2)$, the result-1 in the single pole is unchanged
while the factor 2 contradiction of the result-2 in the double pole is resolved.

Final and reasonable assessment is that we calculate the
following sum when we study the two-loop two-point function
\begin{equation}
I + I_0 + I_{2ct}
\end{equation}
where $I_{2ct}$ is the counter term in the two-loop order($g^4$).
This counter term is chosen to make the sum finite, so that the incomplete 
cancelation in $I+I_0$ is irrelevant.

\vspace{7mm}
%\newpage

\noindent{\itshape Acknowledgement}\ 
The author acknowledges Dr.F.Yuasa and Dr.T.Ishikawa for their
help to check the analytic results by detailed numerical
computation and also for helpful discussion and encouragement.
The author thanks to Dr.N.Nakazawa and Dr.T.Ueda for enlightening
comments on the material studied in this note.

%-----------------------------------------------
\vspace{8mm}
\noindent{\bfseries \large{Appendix-1}}

One of the standard method to compute multi-loop integrals is Nakanishi formula.\rref{3,4}
In case of two-loop integrals, the process is as follows.
The target is the following integral. $N$ is the numerator and $K$ is the number of propagators.
\begin{equation}
I=\int [d\ell_1][d\ell_2] \frac{N}{D_1D_2\cdots D_K}
\end{equation}

First denominator $D_j$'s are combined into a single denominator using
Feynman parameters. Then, through the sequence of linear transformation
(a momentum shift, a rotation, a scale transformation) of loop momenta $\ell_1, \ell_2$,
the denominator turns to $(\ell_1^2+\ell_2^2-V)^K$.
The numerator is also transformed in this process. Then simple integration gives
\begin{equation}
I= \int dx_1\cdots dx_K \delta(1-\sum_{j=1}^K x_j) \sum_a C_a \frac{F_a}{U^{n/2}V^{K-n-a}}
\end{equation}
where 
\begin{equation}
C_a=(-1)^{K+a} \frac{\Gamma(K-n-a)}{(4\pi)^n} , \quad
V=\sum_{j=1}^K m_j^2 x_j - \frac{W}{U} \ .
\end{equation}
Here function $F_a$ stands for the contribution from a part of numerator which has
$2a$ loop momenta.
The function $U$($W$) is the sum of 2nd(3rd) order monomials of Feynman parameters.

For $I$ in \zu{figtwo} they are as follows:
\begin{equation}
U=(x_1+x_2)(x_3+x_4+x_5)+x_1x_2, \qquad
W= q^2[x_5((x_1+x_2)(x_3+x_4)+x_1x_2)] \ .
\end{equation}
The formulae for $F_a$ are omitted here.

%-----------------------------------------------
\vspace{8mm}
\noindent{\bfseries \large{Appendix-2}}

In order to make this note self-contained,
basic formulae for one-loop computation are summarized in this appendix.

\begin{equation}
\int [d\ell]\frac{(\ell^2)^{\alpha}}{(\ell^2-D)^{\beta}} 
= (-1)^{\alpha+\beta}\frac{(4\pi)^{\ep}}{16\pi^2} 
\frac{\Gamma(2-\ep+\alpha)\Gamma(\beta-\alpha-2+\ep)}{\Gamma(2-\ep)\Gamma(\beta)}
\frac{1}{D^{\beta-\alpha-2+\ep}}
\label{eq:oneform}
\end{equation}
where
\begin{equation}
[d\ell]=\frac{d^n \ell}{(2\pi)^n i} \,.
%\label{eq:}
\end{equation}

\begin{equation}
\frac{1}{A^{\alpha}B^{\beta}}=
\frac{\Gamma(\alpha+\beta)}{\Gamma(\alpha)\Gamma(\beta)}
\int_0^1 du dv\, \delta(1-u-v)
\frac{u^{\alpha-1}v^{\beta-1}}{[Au+Bv]^{\alpha+\beta}} \,.
\label{eq:feynforma}
\end{equation}

\begin{equation}
\frac{1}{A^{\alpha}B^{\beta}C^{\gamma}}=
\frac{\Gamma(\alpha+\beta+\gamma)}{\Gamma(\alpha)\Gamma(\beta)\Gamma(\gamma)}
\int_0^1 du dv dw\, \delta(1-u-v-w)
\frac{u^{\alpha-1}v^{\beta-1}w^{\gamma-1}}{[Au+Bv+Cw]^{\alpha+\beta+\gamma}}\,.
\label{eq:feynformb}
\end{equation}

%-----------------------------------------------
\vspace{8mm}
%\newpage
\noindent{\bfseries \large{References}}

\noindent
1) K.Aoki, Z.Hioki, R.Kawabe, M.Konuma and T.Muta, Suppl.Progr.Theor.Phys., vol.73, \\ \ pp.1-225, (1982)

\noindent
2) T.Ishikawa, N.Nakazawa and Y.Yasui, Phys.Rev., vol.D99, pp.073004-1-12, (2019)

\noindent
3) N.Nakanishi, Progr.Theor.Phys., vol.17, pp.401-418, (1957)

\noindent
4) P.Cvitanovi\'{c} and T.Kinoshita, Phys.Rev., vol.D10, pp.3978-4031,(3 papers) (1974)

\noindent
5) E.de Doncker, F.Yuasa and Y.Kurihara, J.Phys.Conf.Ser., vol.365, 012060(8 pages) (2012)

\noindent
6) E.de Doncker, F.Yuasa, K.Kato, T.Ishikawa, J.Kapenga and O.Olagbemi, 
\\ \ Comput.Phys.Commun., vol.224, pp.164-185 (2018)

\noindent
7) F.Yuasa and T.Ishikawa, private communication

\noindent
8) J.Fleischer, M.Yu.Kalmykov and A.V.Kotikov, Phys.Lett. vol.B462, pp.169-177, (1999) ;
J.Fleischer and M.Yu.Kalmykov, Comput.Phys.commun., vol.128, pp.531-349, (2000)

\noindent
9) J.C.Collins, {\itshape Renormalization}, Cambridge University Press, (1984)

%----------------------------------------------
\end{document}